\documentstyle[aps,twocolumn,prl,psfig]{revtex}

\begin{document}
\draft
\title{Universal conductance fluctuations in three dimensional metallic 
single crystals of  Si}
\author{Arindam Ghosh}
\address{Department of Physics, Indian Institute of Science, 
Bangalore 560 012, India}
\author{A. K. Raychaudhuri}
\address{National Physical Laboratory, K.S. Krishnan Road, New Delhi 110 012, 
India}

\date{\today}

\twocolumn[\hsize\textwidth\columnwidth\hsize\csname 
@twocolumnfalse\endcsname

\maketitle
\begin{abstract}
In this paper we report the measurement of conductance fluctuations
in single crystals of Si made metallic by heavy doping ($n \approx 
2-2.5n_c$, $n_c$ being critical composition at Metal-Insulator transition). 
Since all dimensions ($L$) of the samples are much larger than the
electron phase coherent length $L_\phi$ ($L/L_\phi \sim 10^3$),
our system is truly three dimensional. Temperature and magnetic field
dependence of noise strongly indicate the universal conductance 
fluctuations (UCF) as predominant source of the observed magnitude of 
noise. Conductance fluctuations within a single phase coherent region 
of $L_\phi^3$ was found to be saturated at 
$\langle(\delta G_\phi)^2\rangle \approx (e^2/h)^2$. An
accurate knowledge of the level of disorder, enables us to
calculate the change in conductance $\delta G_1$ due to movement of a
single scatterer as $\delta G_1 \sim e^2/h$, which is $\sim$ 2 orders of
magnitude higher than its theoretically expected value in 3D systems.
\end{abstract}

\pacs{72.70.+m, 72.80.Cw, 72.80.Ng}
]

In disordered systems with strong impurity scattering at low temperatures,
the noise can arise from the mechanism of universal conductance
fluctuations (UCF)~\cite{fls}. The origin of the UCF lies in the quantum
interference of multiply backscattered electrons from the scattering centers
and is intimately related to the mechanism of weak localization~\cite{tvr}. 
Theory of UCF~\cite {fls}, showed that at $T =0$ the electrical conductance 
($G$) of a metallic system, is an extremely sensitive function of its 
impurity configuration and alteration of position of
even a single impurity over a sufficient length scale may
induce a conductance change $\delta G_1 \sim e^2/h$. At finite
temperatures, one considers diffusive electronic motion in
regions bounded by electron phase coherence
length $L_\phi$, within which the interference effects are relevant. Total
conductance change due to motion of a number of scatterers inside 
$L_\phi^d$
is additive (as long as $\delta G_1 \ll e^2/h$) and when the number of such
scatterers is sufficiently large,
UCF noise inside one phase coherent region
is saturated with the total noise power
$\langle(\delta G_\phi)^2\rangle \approx (e^2/h)^2$.
At higher temperatures or in cleaner samples with longer
elastic mean free path of the electrons, the noise due to UCF gives way to
that due to mechanisms like local interference (LI)~\cite{pelz}.

Extensive experimental studies on UCF has been carried out in 1D and 2D
disordered systems like, metallic films of Bi~\cite{birge1},
Ag~\cite{birge3}, C-Cu composites~\cite{weiss3}, Li wires~\cite{birge2}
GaAs/AlGaAs heterostructures~\cite{pdb} 
or silicon inversion layers~\cite{wjs} and 
the existence of UCF has been convincingly established in these systems.
However, the absolute magnitude of noise has always been a very roughly
estimated parameter since the nature of the defect and the 
the level of disorder causing the UCF  were unknown in most of the cases.  
In this
paper we report conductance noise measurement in a completely different
class of ``metal'', namely heavily doped single crystalline Si, where
disorder is primarily introduced as substitutional impurities of P (and B)
to the Si matrix.
These are metallic systems with a  low
carrier density ($n$) ($\approx 10 ^{-3}$ times the
carrier density of a metal) and low conductivity ($\sigma$) ($\approx
10^{-2}$ times the conductivity of bad metals) and are close to the
critical composition for Metal-Insulator (MI) transition.

There are the following compelling reasons to carry out the
experiment on single crystals of Si made metallic by heavy doping:
(1) It is a very well defined  system where the number as well as the nature
of defects
are known and the samples used are extremely well characterized
unlike most of the previous studies. 
This will allow us to quantitatively compare the experiment
to the magnitude of the noise predicted by the theory.
(2) The defects which give rise to the noise are in the bulk of the
solid and being single crystal, issues like defects at the surface
or grain boundary
(which often is the case in polycrystalline metallic thin films) are absent
here, (3) This has been the most extensively studied system in investigations
of weak localization and M-I transition and is often taken as model
solid in which new concepts in theories have been tested.
Investigation of electronic phase induced
fluctuation phenomena and UCF has  thus been a long outstanding need
of the field. In particular, it will be search for UCF in a bulk 3D
system in contrast to past studies which were in 2D or 1D.

Polished, $\langle111\rangle$ - Czochralski grown, P and B doped wafers of
Si with thickness $\approx$ 300 $\mu$m were sized down to a length of 2 mm,
width 0.10-0.15 mm and were thinned down by chemical etching
to a thickness of $\approx$ 30 $\mu$m.
These wafers were used extensively in conductivity studies
earlier~\cite{hol}. We have chosen two systems with similar
concentration of donors (P), but one was compensated with B (Si:P,B),
whereas the other was left uncompensated (Si:P)(see table~\ref{tab1}).
They are in the weak
localization regime with 5.5 $< k_Fl <$ 2.3, where $l$ is the elastic mean 
free path. We have studied several samples
of the same system for our experiment.
Noise, electrical conductivity and magnetoresistance (MR) were
all measured in the same sample to avoid any ambiguity. For noise
measurement (done with a temperature stability $|\Delta T/T| < 0.01\%$)
we used a five probe ac technique
(carrier frequency 377 Hz),
aided by digital signal processing methods~\cite{jsco}.
Sample volume for noise
detection ($\Omega$) $\approx 1.5-2.0\times10^{-12}$ m$^{3}$. The
peak current density was kept at $\leq 10^6$ A/m$^{2}$ (power
dissipation $< 50 \mu$Watt) to avoid heating.
Electrical contacts were made by a specially fabricated thermal
wire bonder using gold wires of diameter $\approx 25 \mu$m with
average separation of electrodes $\approx 200-250 \mu$m.  The
contacts, with a temperature independent resistance $\ll$ 1 ohm, were
ohmic. The Hall coefficient was found to
be essentially temperature independent down to 2 K with the
variation in the whole range being $\leq$ 20\%. This ensured that
issues like carrier freeze out are of no consequence here.

Both the samples are metallic with
$\sigma_{4.2K}/\sigma_{300K} >$ 1 (see inset of figure 1).
At $T < 4.2$ K the correction to
conductivity, $\Delta\sigma(T)$ (=$\sigma(T) - \sigma(T=0)$), is
different in the two samples. The Si:P sample showed a dominant
correction which is from the electron-electron interaction~\cite{tvr,hol} and
$\Delta\sigma(T) \approx mT^{1/2}$ with $m <$ 0 . The
compensated sample Si:P,B,
with  higher disorder showed $\Delta\sigma(T) > 0$ as
expected in a sample with correction to conductivity
coming from the weak localization (WL)~\cite{tvr,hol}.

The phase coherence length $L_\phi$ was determined from
MR ($\Delta\rho/\rho$) measurements.
The data are shown in fig.1. For both
the samples MR is negative at low $H$ due to the WL contribution
and at higher $H$ the interaction contribution dominates.
Observed $\Delta\rho/\rho$ were fitted to an expression of MR
consisting of contributions from weak-localization
(consisting of three field scales due to the inelastic
scattering $H_\phi$, spin-orbit scattering $H_{so}$ and the
spin-flip scattering $H_s$) and
interaction. Given the space considerations we are not giving
the detailed fit procedure which are available elsewhere~\cite{tvr}.
The lines
through the data are the calculated MR. In both the systems we observed
$H_\phi \gg H_{so},H_s$. The
$L_\phi$  as obtained from the MR data using the relation
$H_\phi = \hbar/4eL_\phi^2$ are shown as function of $T$ in fig.2.
$L_\phi \propto T^{-p/2}$ where $p \approx$ 1.0. Even at the lowest
temperature, $L_x/L_\phi \approx 10^3$, where $L_x$ is the smallest
dimension (thickness) of the sample. This clearly shows that our system is
truly 3D.

For all the samples studied, the spectral power was 
$\propto V_{bias}^2$ within experimental accuracy and found to scale
with sample volume as $\propto \Omega^{-\nu}$~\cite{ag1}. Typically,
$\nu \approx$ 1.1-1.3.
This scaling was found to be independent of the surface to
volume ratio and valid till low temperatures. This implies that
predominant contribution to the noise arises from the bulk. This
clean inverse dependence is a very important observation because
almost all the past noise studies done on doped
Si ($n \leq 10^{21}$ m$^{-3}$) the source of noise was the
surface or interface~\cite{weiss2}.

In fig.3 we show the typical data of
relative conductance fluctuations
$N\langle(\delta G)^2\rangle/G^2$  as function of $T$.
The variance shows a minima at around 150 K - 175 K.
In the temperature range
$T >$ 175 K, $\langle(\delta G)^2\rangle$ rises very rapidly
following a near exponential dependence on $T$. For $T <$ 100 K,
$\langle(\delta G)^2\rangle$
rises again as $T$ is reduced as $\langle(\delta G)^2\rangle \propto
T^{-q}$, where $q = 0.53\pm0.03$ for Si:P and $0.57\pm0.03$ for
Si:P,B. Very often the noise is expressed through the normalized value
$\gamma$ defined as:

\begin{equation}
\label{eq1}
\gamma = f^\alpha\,S_v(f)\,(\Omega n)/V^2_{bias}
\end{equation}

\noindent $\alpha \approx$ 1-1.2 for $T \geq$ 10 K. At $T <$ 10 K, the     
spectral dependence of the noise power differs significantly from 
$1/f^\alpha$ form as can be seen in the inset of fig.3. 
For comparison, in both systems at $f =$ 3 Hz, $\gamma \approx$ 
1.5 at $T =$ 300 K. It is $\approx$ 0.3 at the minima at
$T =$ 150 K - 175 K and at $T =$ 2 K it has a value $\approx$ 1.
This is about three orders of magnitude higher
than that seen in conventional thin metallic films. In our
experiment, the absolute value of $\gamma$ is reproducible within
$\sim$ 20\% from run to run and 50\% from sample to sample
of the same system, arising
primarily from uncertainty in $\Omega$. The temperature
dependence of $\langle(\delta G)^2\rangle$
(and hence of $\gamma$) including the minima at $T \approx$ 150 K, is
identical to that seen in films of Bi~\cite{birge1}, Ag~\cite{birge3}
and C-Cu~\cite{weiss3} and is qualitatively different
from that seen in lightly doped Si films on sapphire which are deep within
the insulating side~\cite{weiss2}.
It is amazing that the temperature dependence of
$\gamma$ in these three widely different materials are so identical
although the absolute value of $\gamma$ differ. It is even more spectacular
considering the fact that the physical form of the samples (thin film vs
bulk crystal) are completely different and so is the nature of defect that
may cause the $1/f$ noise in these materials.

In investigating the origin of noise, we note that the effect of weak
localization was evident from $\Delta \sigma(T)$ as well as MR for both the
samples. UCF being a  quantum interference related phenomenon, is expected
to be the likely origin of noise in this case. Below 100 K
the diffusive dynamics of electrons is ensured by $l < L_\phi$.
The increase in noise magnitude below
100 K also indicates strongly towards a UCF dominated noise mechanism.
However, the cleanest signature of UCF is the sensitivity of
$\langle(\delta G)^2\rangle$ to an external
magnetic field $H$.  The zero field noise is
expected to get reduced by a factor of 1/2 at some characteristic field 
scale $H_{c1} \approx {\cal A} (h/e)/L_\phi^2$, where
$\cal{A}$  is a constant of the order of unity~\cite{fls}.
This is due to the breaking of time reversal symmetry as the magnetic 
field introduces an extra random phase to the electron's wave function. 
In systems with weak spin-orbit scattering,
a further drop by another factor of 1/2 is 
expected due to spin-symmetry breaking from  the
Zeeman splitting of the conduction electrons at a characteristic field
$H_{c2} \sim k_BT/g\mu_{imp}$, where
$g$ is the g-factor and $\mu_{imp}$ is the
magnetic moment at an impurity site~\cite{ads}.
In fig.4, we plot the observed
variance $\langle(\delta G)^2\rangle$
as a function of $H$. The first reduction of 1/2
occurs very distinctly at field $H_{c1} \approx$ 2.5 mT for both the 
samples. As the temperature is raised, $H_{c1}$ becomes larger.
At higher magnetic field, the fluctuations
are reduced further by an additional factor of 1/2. We identify this as
$H_{c2}$, which agrees well with the calculated value of $\approx$ 1.4 T.
The distinct dependence of the noise on the magnetic field
is a clear indication that a substantial portion of the low
temperature  noise indeed  arises from the UCF mechanism.

From the observed magnitude of rms fluctuations over the experimental 
bandwidth, we can calculate the average variance, $\langle(\delta  
G_\phi)^2\rangle$,
in a single phase coherent box of volume $L_\phi^3$. 
At length scales larger than $L_\phi$, noise
from different phase coherent boxes
get superposed classically and hence, the observed conductance variance
$\langle(\delta G)^2\rangle$ is related to
$\langle(\delta G_\phi)^2\rangle$ as~\cite{fls},

\begin{equation}
\frac{\langle(\delta G)^2\rangle}{G^2} = \frac{L_\phi^3}{\Omega}\,
   \frac{\langle(\delta G_\phi)^2\rangle}{G_\phi^2}
\end{equation}

\noindent where $G_\phi = \sigma L_\phi$. Thus, at $T \approx$ 2 K,
from the observed
$\langle(\delta G)^2\rangle/G^2 = 1.3\times10^{-13}$ for Si:P and
$3.1\times10^{-13}$ for Si:P,B, we get,
$\langle(\delta G_\phi)^2\rangle^{1/2}
\approx 1.5\times(e^2/h)$ for Si:P and $1.2\times(e^2/h)$ for Si:P,B.
This clearly
shows that the noise magnitude within $L_\phi^3$ in both the samples
are saturated. This saturation of $\langle(\delta G_\phi)^2\rangle$
is also reflected in the temperature dependence of the noise.
Using $G_\phi = \sigma L_\phi$, we observe $N\langle(\delta
G)^2\rangle/G^2 \propto L_\phi \propto T^{-p/2}$, where
$p \approx 1.0$. ($N$ is
the total number of carriers in volume $\Omega$.) This matches very
well with our experimental observation.

One of the most basic feature of the UCF theory is the sensitivity of the 
fluctuations towards the movement of a single scatterer which is referred
to as $\langle(\delta G_1)^2\rangle$~\cite{ads1}. 
We show the important result that within a 
phase coherent volume $L_\phi^3$  even  $\langle(\delta G_1)^2\rangle$
is saturated to the value $(e^2/h)^2$. From the UCF theory one obtains
that~\cite{fls,birge1} $\langle(\delta G_\phi)^2\rangle
= n_s \langle(\delta G_1)^2\rangle \times L_\phi^3 \times(L_T/L_\phi)^2$
where $n_s$ is  the density of ``active'' scatterers and 
$L_T = \sqrt{\hbar D/k_BT}$.
Using $\langle(\delta G_\phi)^2\rangle \approx (e^2/h)^2$ for
both the samples,
we get $n_s\langle(\delta G_1)^2\rangle \approx
2\times10^{23}\times(e^2/h)^2$
/m$^3$ for Si:P and $6\times10^{23}\times(e^2/h)^2$ /m$^3$ for
Si:P,B. For both the samples total dopant concentration,
$n_d \approx 10^{25}$ /m$^3$
(see table~\ref{tab1}). Thus for both the samples
$n_s \approx (0.03 - 0.04) \times n_d$ is all that is needed to
saturate the fluctuations $\langle(\delta G_1)^2\rangle \approx 
(e^2/h)^2$. 

An independent estimate of $n_s$ can be obtained from the analysis of
the high temperature noise ($T >$ 150 K). At this temperature range
$l \leq L_\phi$. Under this condition the likely mechanism of noise
can be the LI mechanism ~\cite{pelz}. From the observed
$\langle (\delta G)^2\rangle$ it is possible to estimate the
fraction of active sites $n_{LI}$ taking part in the noise production
by using the relation~\cite{pelz}

\begin{equation}
N\frac{\langle(\delta G_{LI})^2\rangle}{G^{2}} = 
[n_{Si}\lambda(T)\beta_{c}\delta_{c}]^{2} \, \frac{n_{LI}}{n_{Si}}
\end{equation}

\noindent where $n_{Si} = 5\times 10^{28}$ m$^{-3}$, is the atomic density
of Si, $\beta_c \approx$ 0.25 is the anisotropy parameter, $\delta_c
\approx 4\pi/k_F^2$, the average defect cross-section and $\lambda(T)$ is
the net mean free path. We find that within the observed bandwidth
at 300 K, $n_{LI} \approx 5\times10^{-6} n_{Si} \approx 0.03 \times n_d$. 
As $T$ decreases $n_{LI}$ also decreases and below
100 K $n_{LI}$ is $\approx 0.01 \times n_d$. Although $n_s$ and $n_{LI}$
are not the same, they will be of similar magnitude. Thus extending over 
$\sim 10$ decades of frequency, we obtain an
independent estimate of $n_s \approx 0.03 \times n_d$ which
agrees well with that found in the previous paragraph.

The saturation of $\langle(\delta G_1)^2\rangle$ to $\approx (e^2/h)^2$
is a surprising result in a 3D system. According to UCF
theory, in a sample of side $L$ in $d \geq 2$ dimensions, only a
fraction $(l/L)^{d-2}$ of the Feynman
paths passes through a particular scattering site. Assuming that the
transmission amplitude due to this fraction of paths is statistically
independent from that due to rest of the paths, one would expect,
$\langle(\delta G_1)^2\rangle \sim (e^2/h)^2 (l/L)^{d-2}$. More detailed
calculation shows in 3D in a sample of size $L_\phi$,

\begin{equation}
\langle(\delta G_1)^2\rangle = C\,(e^2/h)^2\,
                               \frac{\alpha(k_F\delta r)}{(k_Fl_e)^2}\,
                               \left(\frac{l}{L_\phi}\right)
\end{equation}

\noindent where $C \approx 4\sqrt{3}\pi$ and $\alpha(x) = 1 - 
\sin^2(x/2)/(x/2)^2 \leq 1.0$. 
$\delta r$ is the length scale over which the atomic
displacement takes place. Using the known value of other parameters
and assuming $\alpha \leq 1$,
we get $\langle(\delta G_1)^2\rangle \leq 0.01\times(e^2/h)^2$
in Si:P and $\leq 0.03\times(e^2/h)^2$ in Si:P,B. This
value of $\langle(\delta G_1)^2\rangle$ is nearly two orders of magnitude
less than the saturation value of $(e^2/h)^2$ observed experimentally.

Thus UCF theory underestimates  $\langle(\delta G_1)^2\rangle$
for 3D systems.  Our result indicates that the assumption of the
theory of uncorrelated (statistically independent) Feynman paths
for 3D system may not be the correct assumption. From the experimental
observation of the saturation of noise, we have to reach the conclusion
that when a
scattering center in the system moves, transmission probability associated
to even those paths which are not passing through the center gets
affected by correlation effects. It would be interesting to know whether
this correlation
among the Feynman paths is due to electronic phase coherence intrinsic to
any 3D system or a finite correlation length effect in systems near
MI transition.

In conclusion, we have observed saturated
universal conductance fluctuations in a 3D disordered system, namely
single crystals of silicon made metallic by heavy doping. We find that the
existing theory of UCF grossly underestimates
the absolute magnitude of noise in
such systems even though the qualitative features do show satisfactory
agreement~\cite{hc1}.

We would like to thank Prof. D. Holcomb of Cornell University for the
Si samples and Prof. T.V. Ramakrishnan for a number of helpful
discussions.

\begin{table}\caption{}
\label{tab1}
\begin{tabular}{|l|c|c|c|c|c|c|}
Sample & $\sigma(0)$ (S/m) & $n_d$ (m$^{-3}$)& K   & $l$ (nm) & 
$\Omega$ (m$^3$)  & $L_T$ (nm) \\ \hline\hline            
Si:P   & 3.1$\times$10$^4$ & 1$\times10^{25}$& 0.0 & 8.2& 1.9$\times10^{-12}$ & 8.7 \\\hline
Si:P,B & 1.2$\times$10$^4$ & 1$\times10^{25}$& 0.4 & 3.4  & 1.5$\times10^{-12}$ & 6.9\\
\end{tabular}
\end{table}

{\bf\large Figure caption:}

{\bf figure 1:} Magnetoresistance (MR) as a function of magnetic field ($H$).
The inset shows temperature variation of the zero field conductivity. 
The solid
lines are fit to the MR data according to the weak localization and
electron-electron interaction correction.

{\bf figure 2:} The temperature dependence of the phase breaking length 
$L_\phi$ determined from magnetoresistance fits. The solid lines show, 
$L_\phi \propto T^{-1/2}$ in this temperature range.

{\bf figure 3:} Conductance fluctuations as a function of temperature. The 
inset shows the spectral dependence of noise power at the lowest $T$.

{\bf figure 4:} The magnetic field dependence of the conductance 
fluctuations.

\begin{references}
\bibitem{hc1} We also find that $H_{c1}$ obtained
in noise experiments is much less than the estimate $H_{c1} \approx
(h/e)/L_\phi^2$. Further investigations on this issue will be published in
a future publication.
\end{references}
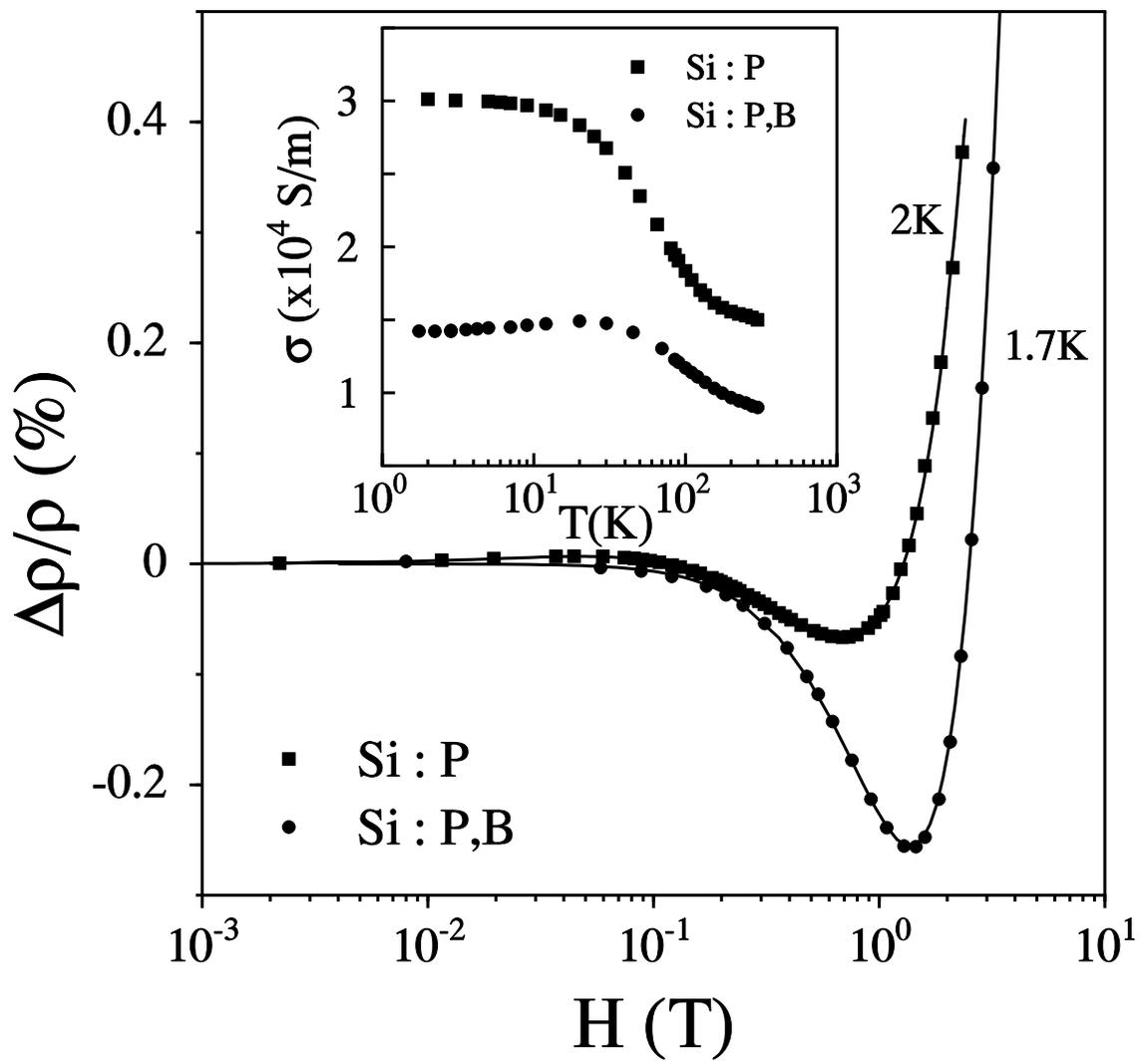
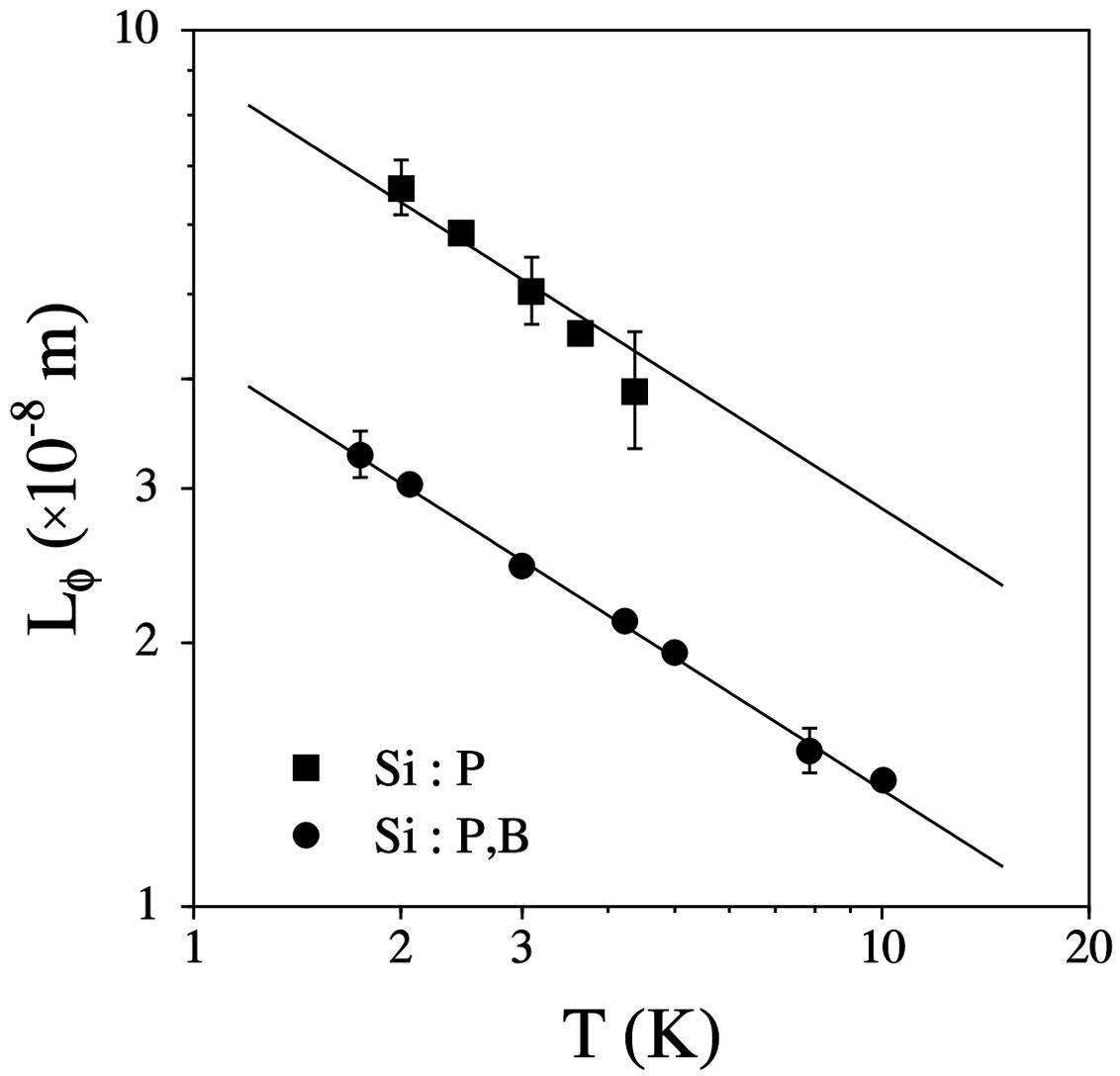
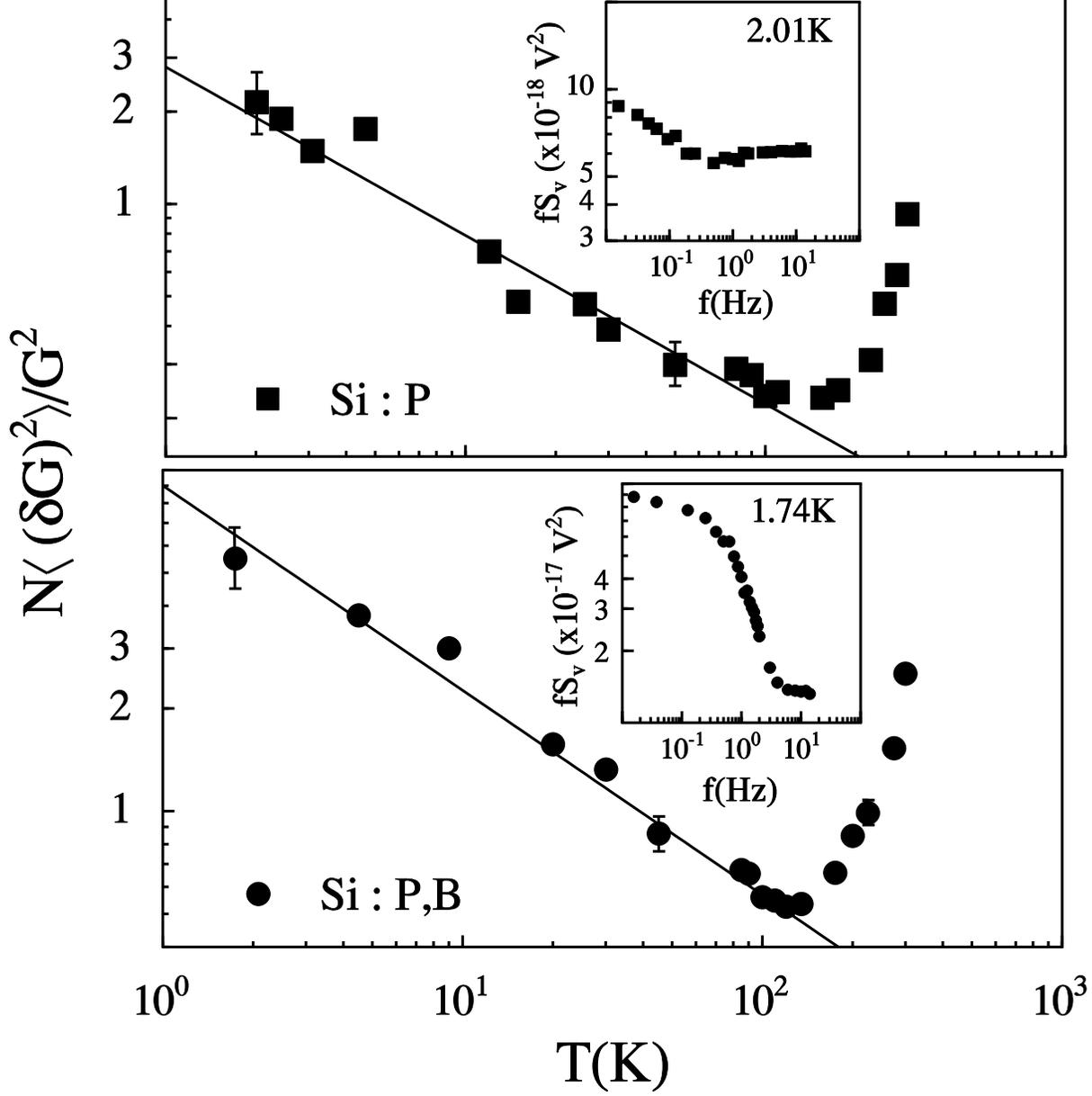
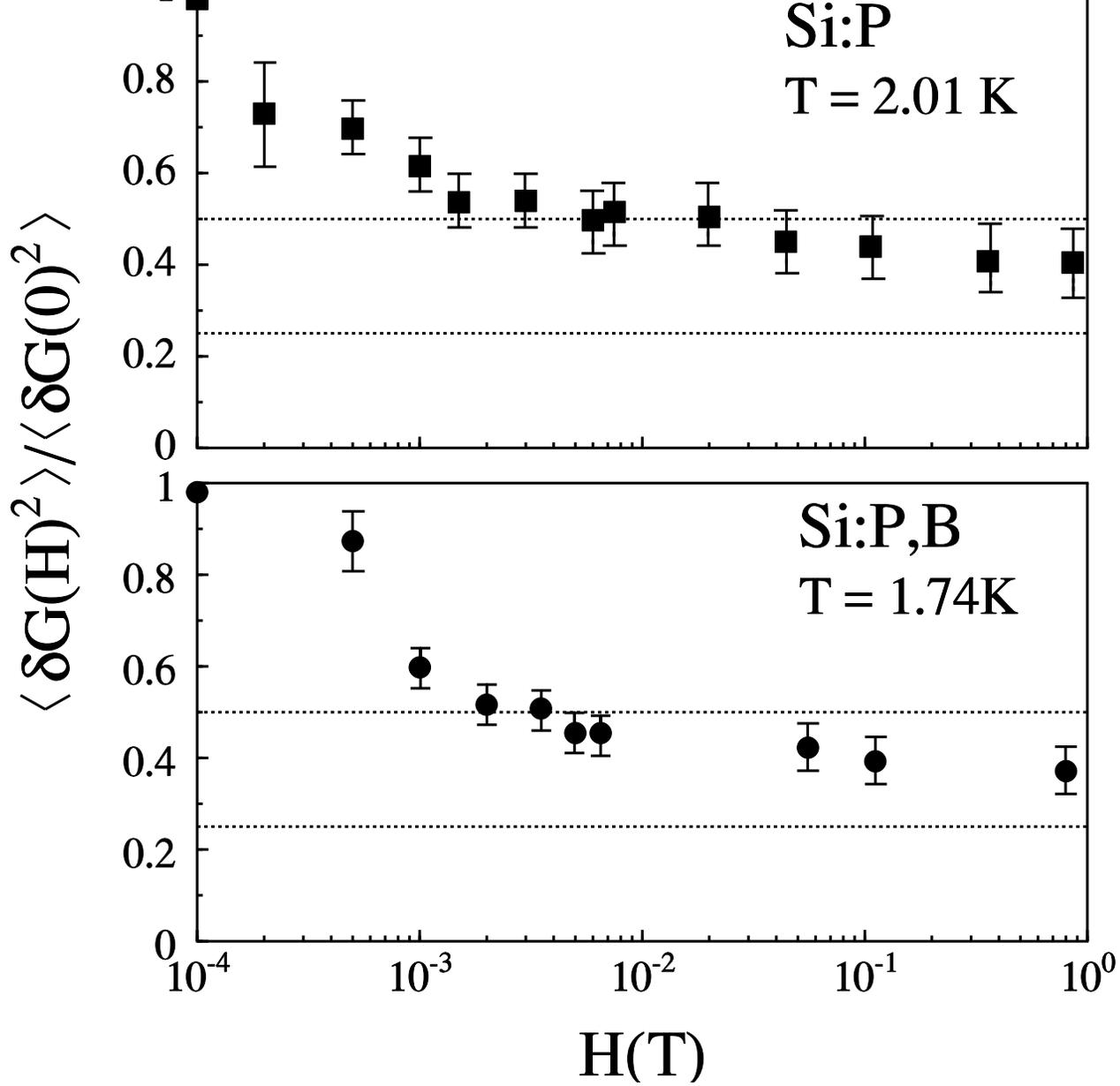
\end{document}